\documentclass{style/nseJournal}

\begin{document}

\title{Modelling the Performance of Tritium Process Monitors from First Principles} 

\addAuthor{\correspondingAuthor{Nicolas J. Sovare}}{a}
\correspondingEmail{njs1893@rit.edu}
\addAuthor{Walter T. Shmayda}{b}

\addAffiliation{a}{Rochester Institute of Technology}
\addAffiliation{b}{Tritium Solutions Inc.}

\addKeyword{In-line process monitors}
\addKeyword{Tritium systems}
\addKeyword{Computational models}

\titlePage

\begin{abstract}
Ionization chamber-based, in-line tritium process monitors play an important part in determining the behavior of a tritium system. The one-liter detection volume monitor has been well-characterized through experiment and shown to respond linearly to tritium concentrations for the range of 1 $\mu Ci/m^3$ to 1 $Ci/m^3$. Additionally, it has been shown to behave nonlinearly for low voltage on the central anode and at low gas pressures. A computational model was developed from first principles that successfully describes the behavior of the one-liter monitor for each of these regimes. Predictions from the model are compared to previously-collected experimental data in order to determine validity and to tune the model. The model will be applied to novel detector geometries and designs.
\end{abstract}

\section{Introduction}

The one liter detection volume, ionization chamber-based process monitor contains an anode held to a negative voltage within a well-defined detection volume. This produces an electric field that permeates the detection region. As it is attached in-line to a tritium process, tritium is directly delivered into and out of the chamber via a carrier gas. While tritium is inside of the chamber, it has a chance to undergo beta decay, producing a helium ion and an energetic beta particle. Under ideal conditions, this beta particle deposits all of its energy into the surrounding carrier gas, producing several additional ions from the gas. The permeating electric field draws these ions to the central anode, where they will recombine with electrons from the anode, producing neutral particles. The draw of electrons from the anode is detected as current.

Under ideal circumstances, this produced current is equal to the "saturation current~\cite{ProcessMonitors}", $I_{sat}$, which is linear with the concentration of tritium in the process:

\begin{equation}
    I_s(fA) = \frac{E(eV)\  c(Bq/m^3) \ V(m^3) }{W(eV)} \cdot 1.6\times10^{-4}\left(
\frac{fC}{ion}\right)
    \label{eqn:SaturationCurrent}
\end{equation}

\noindent Here, E is the average energy of the electron emitted via beta decay (5.7 keV), c is the concentration of activity, V is the detection volume of the process monitor, and W is the average energy required to produce an ion-pair inside of the carrier gas. 

However, eq.~(\ref{eqn:SaturationCurrent}) fails to predict the true current seen by the detector under non-ideal circumstances. At low gas pressure, the range of emitted beta particles increases to the point that they begin to reach the outer wall of the chamber while retaining most or all of their energy. This has the effect of decreasing pair production, and causes a significant decrease in output current. In addition, reducing the magnitude of the voltage on the central anode decreases the speed at which ions drift towards the center of the detector. At sufficiently low voltage, ions will accumulate in the detection region, increasing the probability of recombination with free electrons in the chamber. Upon recombination, a neutral particle is produced, which cannot be detected by the monitor. This also has the effect of reducing current seen by the detector. 

Motivating the development of a program is the fact that these non-ideal effects don't lend themselves to the development of a simple equation to predict the output current, and that future designs for process monitors include more complicated geometries that are less likely to be characterizable by eq.~(\ref{eqn:SaturationCurrent}). This paper underlines the methods employed by the model to make accurate predictions for the current seen by the one-liter process monitor, as the first step in guiding the design of process monitors for high activity applications.

\section{Program Implementation}
\label{sec:equations}

\subsection{Preparing the Simulation}

The primary mover of ions in the chamber is the electric field produced from the difference in potential between the central anode and the exterior walls. The voltage of the central anode and walls will not change during a given trial, so this can be calculated ahead of time and referenced later. In order to determine the electric field, we first discretize the volume into small bins so that we can use finite-difference methods to define the potential and subsequent electric fields. Next, we hold constant the voltages of the bins comprising the outer wall and the anode. From here, the Jacobi method of iteration is used to solve the Laplace equation for the voltage everywhere inside of the chamber. Once obtained, numerical differentiation is used to obtain the electric field throughout the region. The results of these calculations are shown in fig.~(\ref{fig:AnodeVoltage}):
\begin{figure}[h]
    \centering
    \includegraphics[width=0.75\linewidth]{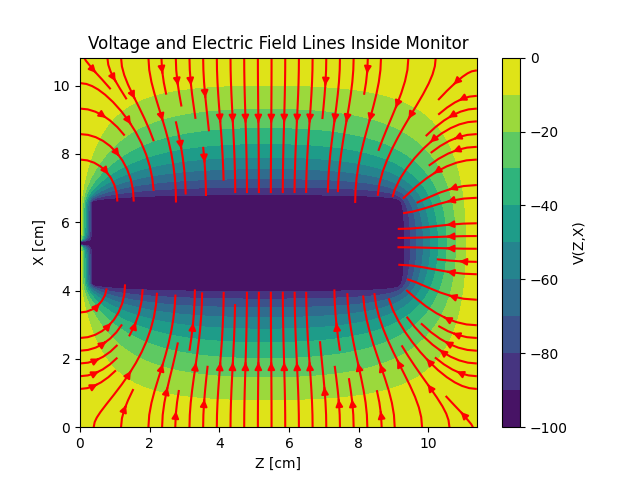}
    \caption{A radial cross-section of the electric potential (in volts) as a function of position in the chamber. Overlaid on the graph is the direction (but not the magnitude) of the electric field, which lies perpendicular to the equipotential lines.}
    \label{fig:AnodeVoltage}
\end{figure}

To initialize the model for a trial, an electric field is selected to match the desired anode voltage. Additionally, the pressure of the carrier gas and the true concentration of activity are decided at this time. The concentration of activity is used alongside the detection volume to determine the rate of beta decays in the trial. The number of decays per second in the chamber is the product of the concentration and the volume. The carrier gas pressure is used to precompute the range of emitted beta particles as well as the distribution of ion deposition resulting from their interactions with neutral particles. Both of these features can be found from eq.~(\ref{eqn:BetaRange}):

\begin{equation}
    r(p) = \frac{kRT}{PZ}E^{1.265 - 0.094ln(E)}
    \label{eqn:BetaRange}
\end{equation}

This is an empirical equation, where, r is the average distance in cm traveled by an electron of initial energy E (MeV) through a medium of atomic number Z, and pressure P (Torr). R is the molar gas constant and k is an empirically determined constant.

The program takes, as input, an estimate for the average Debye length inside of the detection region. The Debye length~\cite{PlasmaPhysicsBook}, $\lambda_D$, describes the effective range of local electric effects in the plasma and is given as:

\begin{equation}
    \lambda_D = \sqrt{\frac{\epsilon k_BT}{Nq^2}}
    \label{eqn:DebyeLength}
\end{equation}

Here, $\epsilon$ is the permittivity. $k_B$ is Boltzmann's constant, T is the average ion temperature, N is the ion density, and q is the ionic charge. Beyond the distance, $\lambda_D$, from a single ion, its Coulombic effects are screened out by other charges in the plasma. As a result, the Debye length serves as an important characteristic for determining an effective range to be considered when handling electron-ion recombination as well as ion-ion interactions. Throughout the evolution of the program, statistics are collected in order to determine the actual Debye length. After the simulation is complete, this information is used to determine the validity of the simulation. 

\subsection{Evolving the System in Time}

With the base parameters set, the system can be evolved in time. To begin, a timestep is determined, over which the system is evolved. In practice, a step size of 0.01 seconds has proven sufficiently small to avoid incurring numerical errors during the simulation. In each timestep, the program follows a five part process to accurately handle the creation, interaction, recombination, evolution, and detection of ion-pairs inside of the detector.

\subsubsection{Ion Deposition}
Operating on the assumption that tritium is evenly distributed in the carrier, and thus evenly distributed inside of the ionization chamber, the exact locations of tritium decay are selected from a uniform distribution within the entire detection region. These positions are stored with 32-bit floating point precision and used as a starting point for beta-decay. At this point, the range determined in eq.~(\ref{eqn:BetaRange}) is used, along with a random choice of direction to attempt to place ions in a straight line away from the location of the decay event. The exact positions of ion placement on the line vary based on the rate of energy loss along the path. If the program attempts to place an ion outside of the detector walls, it is removed and will not be counted. This allows the program to simulate the loss of energy from beta particles reaching the walls without depositing all of their energy into the gas.

Currently, the program assumes that the energy associated with each beta particle is the average value of 5.7 keV. In reality, the energy of each beta particle will be different and could range up to 18.6 keV. For most carrier gas pressures, this difference doesn't matter, but it is likely responsible for the deviation from experiment seen in fig.~(\ref{fig:PressureDependence}). 

\subsubsection{Ion-Ion Interaction}
In addition to the electric field between the central anode and the outer walls, which is the primary mover of particles inside of the chamber, there is also a secondary effect that results from the interactions between the individual ions and electrons, which all produce their own electric fields. Solving Coulomb's law for the pairwise interactions between each ion and electron inside of the detection region quickly becomes untenable, suggesting that an alternative method be used. The solution comes in the form of the Particle-In-Cell~\cite{PlasmaPhysicsBook} method (PIC), which works by discretizing the volume of the detector, determining an appropriate charge density in each volume element (or cell), and solving the Poisson equation using the same methods as described above for solving the Laplace equation. 

The appropriate charge density in each cell is determined by the use of a "weighting function", which determines the contribution from each particle to the overall density in the cell from the position of the particle. After the Poisson equation is solved, the electric field is carried back to each particle by moving backwards through the weighting function. More complex and accurate weighting functions consider the distance from the particles to the centers of the closest set of neighboring cells when calculating the charge density in each cell. However, it was determined that for the purposes of this work, the ion-ion interactions have a minor enough contribution that the error incurred by using the simpler "zeroth order" weighting function (The charge density of each cell is equal to the net charge in the cell divided by the volume of the cell) is negligible and well worth the benefits of reducing computational cost.

When determining the appropriate volume of each cell, the only important characteristic is the Debye length, as that is the range over which local interactions can be seen. In order to avoid loss of information, the PIC method must be able to reliably resolve this length. As a result, the length of each bin is chosen to be one half of the previously calculated Debye length, so that the Debye length is considered across the entire detection volume.

\subsubsection{Ion-Electron Recombination}
As mentioned previously, the longer that ions and electrons are allowed to accumulate in the detection region without reaching the central anode, the more likely they are to being recombining and producing neutral particles. By manipulating the recombination equation discussed in (Ref.~\cite{VonEngel})  the rate of recombination is expressible as a function of the number of electrons and ions in a given volume, expressed as:

\begin{equation}
    -\frac{dn_i}{dt} = \rho_i\frac{n_in_e}{V}
    \label{eqn:recombination}
\end{equation}

\noindent where $n_i$ and $n_e$ are the number of ions and electrons respectively, V is the volume of the region being considered, and $\rho_i$ is the recombination coefficient for the particular ion. In order to apply this to the model, we transform eqn.~(\ref{eqn:recombination}) into a difference equation that can be implemented in each timestep:

\begin{equation}
    \Delta n_i = -\rho\frac{n_in_e}{V}\Delta t
    \label{eqn:discreteRecombination}
\end{equation}

To handle recombination, the detection region is divided into volumes of length equal to the Debye length. This is because screening should prevent recombination from occurring beyond that length. On each timestep, this equation is applied across each of the volumes to determine how many ions need to be removed from the simulation. As an ion being removed requires it to recombine with an electron, an electron is also removed at this point, maintaining an overall neutral plasma. At this point in the simulation, information is being collected on the ion density inside of the detection region, which makes this a good area to determine if the estimated Debye length fits with the one found by the program. If the whole trial completes and it is found that the original estimation differs from the actual Debye length found by the program, the trial will be restarted using the new Debye length as a new estimate. 

\subsubsection{Position Evolution}
With all of the interactions handled for the given timestep, the program moves on to resolving the effects of those interactions. The program calculates the electric field acting on each particle, then moves particles accordingly. In an ideal gas with no collisions, this would be done by calculating the force due to the electric field and tracking accelerations. However, this is not the case for the plasma inside of the chamber. Ions, traveling under the influence of their net electric fields, will attempt to accelerate, but will quickly collide with neutral members of the carrier gas. As a result, a sort of drag force is applied to them, and an average velocity is quickly reached. The average velocity acquired by an ion in a neutral gas per unit electric field is known as the mobility~\cite{VonEngel}, $\mu$, where velocity is given as:

\begin{equation}
    v = \mu E/p
    \label{eqn:mobility}
\end{equation}

\noindent where E is the electric field and p is the pressure of the gas. This equation is used to find the velocity of each charged particle in the simulation on a given timestep. Then, with the velocity of a given particle, a simple Euler-step method is used to evolve the positions over one timestep. Euler's method is chosen here because at the distances and velocities considered, its introduction of noise is minimal for the given timestep size. Using a more advanced method would allow for a larger timestep if evolving positions was all that is required, but the limiting features of this process are actually the ion-ion interactions and the ion-electron recombination, which require the shorter timestep in order to avoid incurring error.

\subsubsection{Ion Detection and Removal}
On each timestep, positions of ions are checked to see if they have reached the central anode. A line is drawn between the previous position and the current one to determine if ions have moved through a piece of the detector during the move step. If either of these events have occurred, the ion is removed from the simulation and its charge is added to rolling sum that goes over all timesteps. The average current seen by the detector is calculated by dividing this sum by the total elapsed time. 

\begin{figure}[h]
    \centering
    \includegraphics[width=0.75\linewidth]{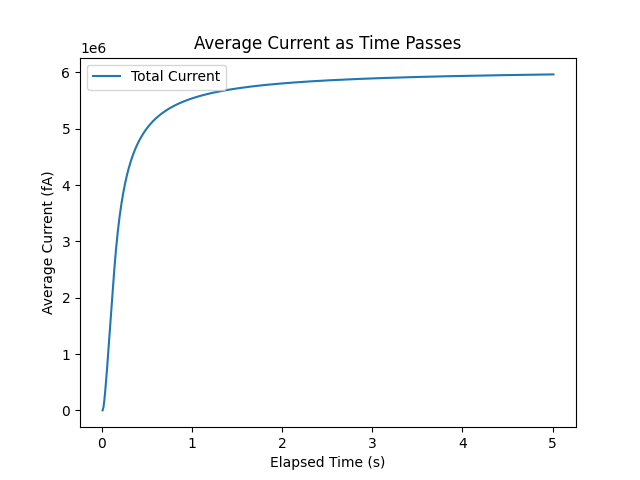}
    \caption{Average current seen by the chamber throughout a series of timesteps. Within five seconds, the average current calculated by the model converges to a stable value, indicating to the program that no further iterations are required.}
    \label{fig:ConvergenceOfCurrent}
\end{figure}

The average current seen at this timestep is compared to the current seen at previous timesteps. If the value has stabilized within a preset range of error, typically one or two percent, then the simulation can be ended early as seen in figure~\ref{fig:ConvergenceOfCurrent}. Typically, at high anode voltages, this occurs within five seconds of simulated time. If the Debye length was found remain unchanged during the recombination step, then this data is saved. Otherwise, the trial is restarted with an improved estimate for the Debye length. By modifying parameters across multiple runs of this procedure, such as anode voltage, carrier gas pressure, and carrier gas species, the model can make predictions on how the current seen by the process monitor responds to those variations.

\section{Results}

\begin{figure}[h]
    \centering
    \includegraphics[width=0.75\linewidth]{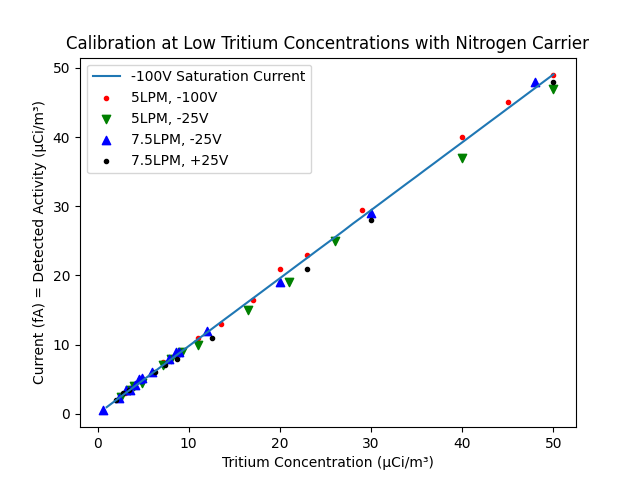}
    \caption{At 50 evenly spaced concentrations between 1 and 50 $\mu Ci/m3$, the simulation is run for a fixed gas pressure of 760 Torr and anode voltage of -100 V.  The carrier gas used for this series of trials was nitrogen. The blue line is the predicted response of the process monitor according to the model. The points correspond to previous experimental data~\cite{ProcessMonitors} collected under ideal conditions.}
    \label{fig:LowActivitySaturation}
\end{figure}

\begin{figure}[h]
    \centering
    \includegraphics[width=0.75\linewidth]{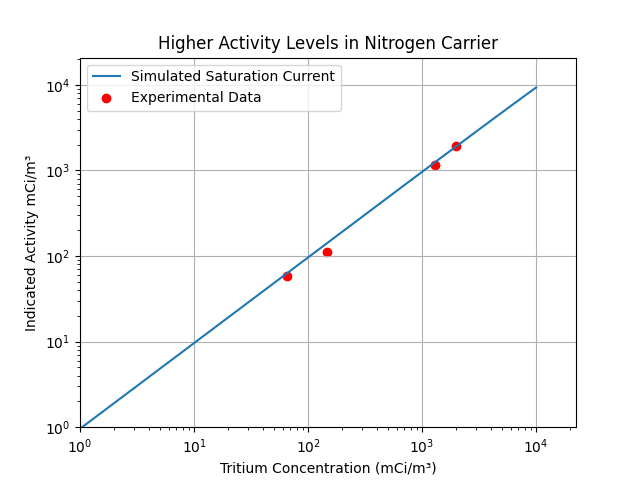}
    \caption{Another 15 evenly spaced concentrations are selected between 1$\mu Ci/m3$ and 104 $mCi/m3$. The same conditions of -100V and 760 Torr nitrogen are used for this trial. Again, the blue line is the prediction of our model, while the red points correspond to previously measured behavior of the process monitor~\cite{ProcessMonitors}. The response of the detector remains in the ideal regime as the electric field remains sufficient to 
keep ion densities low at higher tritium concentrations.}
    \label{fig:HighActivitySaturation}
\end{figure}

Initial simulations were carried out in the saturation regime over two ranges of tritium concentration in a nitrogen carrier gas. First at a lower concentration of 1-50 $\mu Ci/m3$, then from 1-104 $mCi/m3$. These results of these trials are plotted against previous, experimental results~\cite{ProcessMonitors} for comparison. As expected according to previous experimental data, the predicted response from the process monitor in figure~\ref{fig:LowActivitySaturation} and figure~\ref{fig:HighActivitySaturation} remains linear to the concentration of activity in the chamber and there is a one to one correspondence between predicted and measured values.

\newpage

\begin{figure}[h]
    \centering
    \includegraphics[width=0.75\linewidth]{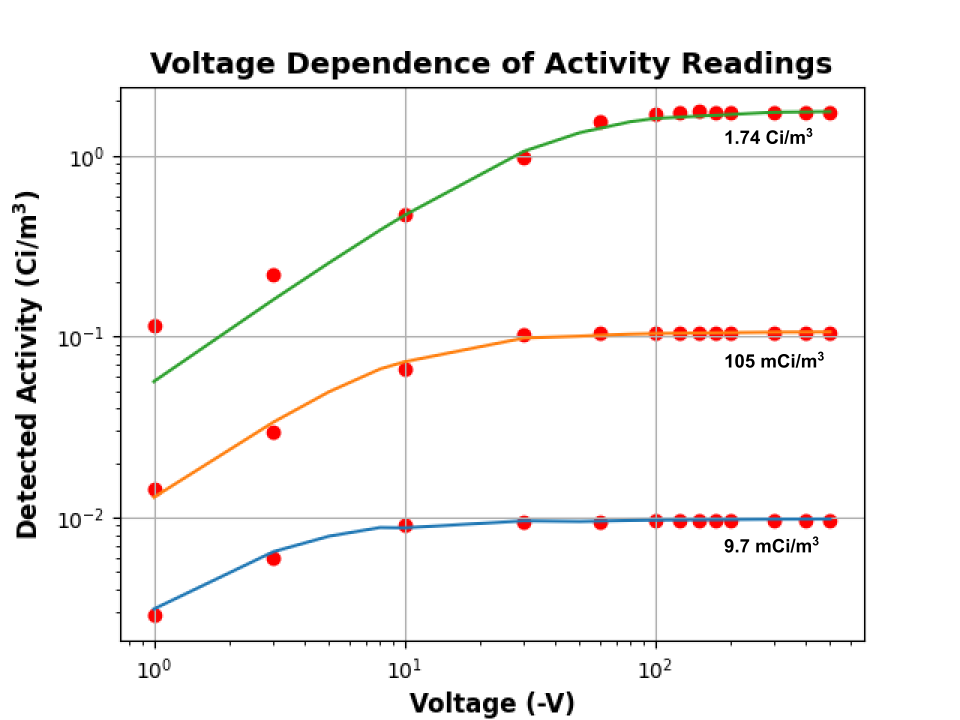}
    \caption{At -1, -3, -5, -8, -10, -30, -50, -80, -100, -300, and -500 Volts on the central anode, the simulation is run for a fixed argon carrier gas pressure of 760 Torr at the three fixed concentrations of activity: 9.7 $mCi/m^3$, 105 $mCi/m^3$, and 1.74 $Ci/m^3$. The selected recombination coefficient, $\rho = 3\times10^{-6} \frac{cm^3}{ion \cdot sec}$, was selected to align the model to experimental data. The red points represent previously collected experimental data~\cite{ProcessMonitors}, while the colored curves correspond to the predictions made by the program. As with experiment, it is found that the magnitude of the voltage required to remain in the ideal regime increases with the true concentration of activity inside of the chamber.}
    \label{fig:VoltageDependence}
\end{figure}

Next, the ability of the model to handle variations of anode voltage was tested. At high anode voltages, the ions generated inside of the process monitor are quickly delivered to the central anode before they can be lost to recombinative effects. As the voltage of the central anode decreases, ion velocities drop, resulting in higher densities of ions within the volume. In agreement with previous experiments~\cite{ProcessMonitors}, we find in figure~\ref{fig:VoltageDependence} that below a certain threshold voltage for the anode, the current seen by the anode will drop below ideal behavior and instead follow a power-law relationship, resulting from increased recombination rates. The deviation between experiment and prediction in the 1.74 $Ci/m^3$ case at low anode voltages is attributed to surface contamination of the inner surfaces of the detector.

\begin{figure}[h]
    \centering
    \includegraphics[width=0.75\linewidth]{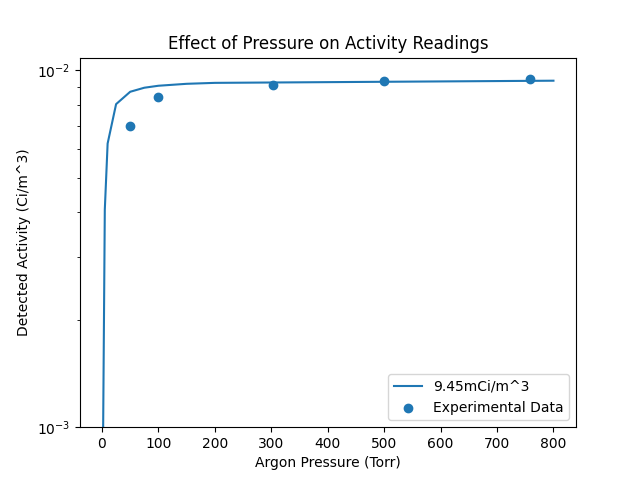}
    \caption{The model was tested at a fixed concentration of 9.45$mCi/m^3$, with anode voltage at -100V, and an argon carrier gas. Detected activity is measured as a function of carrier gas pressure. The points are previously collected experimental data~\cite{ProcessMonitors}, while the line is the prediction curve generated by the model. In agreement with experimental data, we find that there is little apparent dependence on pressure above 100 Torr. However, output current drops precipitously once pressure falls below 70 Torr and the range of the beta particles becomes great enough to deposit a significant amount of energy into the outer walls.}
    \label{fig:PressureDependence}
\end{figure}

\newpage Finally, the model was tested for its ability to handle pressure-dependence. At high pressures, the beta particles produced by decay will have a sufficiently small range to deposit all of their energy into the gas and become thermalized. However, as pressure decreases, the range of the beta particles increases. As this range increases and grows comparable to the size of the chamber, betas begin to reach the outer wall of the detector before all of their energy can be released into the gas. That energy is then lost from the system, resulting in fewer ion-pairs being generated. This behavior is recreated in figure~\ref{fig:PressureDependence}.

\newpage
\section{Discussion and Conclusions}
While the model is highly accurate for a range of conditions, there are deviations from experimental data when considering the non-ideal regime of trials. Prediction lines in figure~\ref{fig:VoltageDependence} agree with experiment at high anode voltages, but show minor nonlinear effects as the magnitude of anode voltage drops. In addition, the determined recombination coefficient for argon in this simulation was $3\times10^{-6} \frac{cm^3}{ion \cdot sec}$, which is an order of magnitude higher than the value $3\times10^{-7} \frac{cm^3}{ion \cdot sec}$ published in \cite{Recombination}. This suggests the possibility of further improvements to the recombination model.

Figure~\ref{fig:PressureDependence} also shows minor deviations between experimental data and the prediction produced by the model. As mentioned previously, this likely arises from the choice to give each decay electron an average energy instead of selecting electron energies from the allowed distribution for tritium decay. If the distribution were used instead, it would follow that more of the generation of ion-pairs would result from the higher energy beta particles, which would start losing their energy to the outer walls a bit sooner as a result of their greater range. This would have the likely effect of shifting the prediction curve down around the 100 - 200 Torr range.

A model has been developed from first principles, which accurately models the physics inside of the 1L Tritium Monitor. The program has been found to reliably reproduce experimental data on the behavior of the process monitor, both in ideal and in non-ideal experimental regimes. With the model describing the physics inside of the detector, the model can now be expanded to predict different geometries, such as the 20cc wire cage monitor design. This model will be useful to predict and inform the design of newer process monitors.

\section*{Acknowledgments}
This research was funded in part by Torion Plasma Corporation and by Torion USA Inc. The authors acknowledge their contribution.

\bibliographystyle{style/ans_js}                                                                           
\bibliography{bibliography}

@book{PlasmaPhysicsBook,
	Author = {Birdsall, C.K. and Langdon, A.B.},
	Edition = {1st},
	Publisher = {IOP Publishing Ltd.},
	Title = {Plasma Physics via Computer Simulation},
	Year = {1991}}

@article{ProcessMonitors,
	Author = {Kherani, N.P. and Shmayda, W.T.},
	Journal = {Fusion Technology},
	Number = {},
	Pages = {340--345},
	Title = {In-Line Process Tritium Monitors},
	Volume = {21},
	Year = {1992}}

@book{VonEngel,
        Author = {von Engel, A},
        Edition = {1st},
        Publisher = {Oxford at the Clarendon Press},
        Title = {Ionized Gases},
        Year = {1955}}

@article{Recombination,
        Author = {Biondi, M.A. and Brown, S.C.},
        Journal = {Physical Review},
        Number = {11},
        Pages = {1697-1700},
        Title = {Measurment of electron-ion recombination},
        Volume = {76},
        Year = {1949}}

\end{document}